\documentclass{ws-mpla}

\begin{document}

\markboth{Lian liu, Xuan-min Cao, Hui Liu}
{static dielectric function and potential from AdS/CFT}

\catchline{}{}{}{}{}

\title{THE STATIC DIELECTRIC FUNCTION AND INTERACTION POTENTIAL IN STRONG COUPLING WITH AdS/CFT}

\author{Lian Liu}
\address{Physics Department, Jinan University, Guangzhou 510632, China}
\author{Xuan-min Cao}
\address{Physics Department, Jinan University, Guangzhou 510632, China}
\author{Hui Liu}
\address{Physics Department, Jinan University, Guangzhou 510632, China\\tliuhui@jnu.edu.cn}

\maketitle
\pub{Received (Day Month Year)}{Revised (Day Month Year)}

\begin{abstract}
In this paper, we studied the static dielectric function and interaction potential in strong coupling limit with AdS/CFT correspondence. The dielectric function is depressed compared with that in weak coupling. The interaction potential then presents a weaker screening characteristics in strong coupling, which indicates a smaller Debye mass compared with weak coupling.

\end{abstract}

\ccode{PACS Nos.: 11.10.Wx, 12.38.Aw, 12.38.Mh.}

\section{Introduction}
Color confinement is a crucial feature of quantum chromodynamics (QCD). Because of the confinement, one could not observe free quarks and gluons by dynamical method but would expect to  observe them by creating a thermodynamic environment such as that in relativistic heavy ion collision. In the relativistic heavy ion collision, an extremely high temperature and density condition would induce a deconfined phase transition that makes the nuclear matter transform to the free quark-gluon plasma (QGP) phase. However, the QGP signals, which we have discovered from recent experiments, indicate that the quark matter created in the Au-Au 200GeV collision is strongly coupled rather than weakly coupled as expected before\cite{1,2,3,C}. Therefore understanding the properties of strongly coupled matter becomes an interesting issue.

Since perturbation expansion fails in the discussion of strong coupling, AdS/CFT correspondence is currently developed to discuss  strongly coupled QGP (sQGP). This  correspondence builds a connection between strong coupling limit in conformal field theories (CFT) and weak coupling limit in string theories or supergravity on certain background. Rather, the correspondence between the $\mathcal N=4$ supersymmetric Yang-Mills (SYM) theory and the type IIB string theory in $AdS_5\times S^5$ space allows us to obtain the properties of strong coupling in quantum field theory by calculating those counterparts of weak coupling in classic gravity theory \cite{A,4,5,6,7}.

In various aspects of medium properties, dielectric function and interaction potential might be the most fundamental ones. The dielectric function reflects the medium response to an external source, which actually  embodies the difference between fields in medium and in vacuum. The interaction potential describes the effective dynamics of particles considering the modification of nearby particles, which determines the macroscopic behavior of the system through thermodynamics.

We calculate the static dielectric function and interaction potential in strong coupling limit with AdS/CFT correspondence in this paper, comparing them with those in weak coupling limit. The whole paper is arranged as follows. Section 2 is the formalism of static dielectric function and static interaction potential in the linear response theory. In the section 3, we review the static equation of motion (EOM) from AdS/CFT in R-charge model. In section 4, we work out the numerical solutions of the EOM and present the static dielectric function and interaction potential, then compare them with the weak coupling ones. The last section is conclusion. The momentum notation in this paper follows $Q_{\mu}=(q_0,\mathbf{q})$ and $Q^2=Q_\mu Q^\mu=q_0^2-\mathbf{q}^2$.

\section{Linear response framework}
In electromagnetism, the electric displacement field \textbf{D} represents how an electric field \textbf{E} influences the organization of electrical charges in a given medium, including charge migration and electric dipole reorientation. From the macroscopic point of view, when it comes to a linear, homogeneous, isotropic medium, the medium field \textbf{D} and the vacuum field \textbf{E} are related by a factor in the simple case  which says $ \mathbf{D}(Q)=\varepsilon(Q)\mathbf{E}(Q) $, where the so-called dielectric function $\varepsilon(Q)$ is the function of frequency and momentum. In other words, the macroscopic medium effect is reflected by the dielectric function. From the microscopic point of view, the medium response to an external field is determined by the polarization process, which means the polarization tensor governs the medium effect namely the dielectric function.

For a massless vector field, the action in medium is
\begin{equation}\label{action}
\Gamma=\Gamma_0+\Gamma_1,
\end{equation}
where
\begin{equation} \label{action0}
\Gamma_0=-\frac{1}{2}\int \frac{d^4 Q}{(2\pi)^4}(\mathbf{E}^2-\mathbf{B}^2)
\end{equation}
is the action in vacuum.
In the linear approximation, the correction $\Gamma_1$ is expressed by the polarization tensor $\Pi_{\mu\nu}$ as
\begin{equation} \label{action1}
\Gamma_1=-\frac{1}{2}\int \frac{d^4Q}{(2\pi)^4}A^\mu(-Q)\Pi_{\mu\nu}(Q)A^\nu(Q),
\end{equation}
where $A_{\mu}$ is the electromagnetic vector field.

We can decompose the tensor by employing projection operators
\begin{eqnarray*}
P^T_{00}&=&P^T_{i0}=P^T_{0i}=0,\\
P^T_{ij}&=&\delta_{ij}-\frac{q_{i}q_{j}}{q^2},\\
P^L_{\mu\nu}&=&\frac{Q_{\mu}Q_{\nu}}{Q^2}-g_{\mu\nu}-P^T_{\mu\nu}.
\end{eqnarray*}
With these projection operators, one rewrites the polarization tensor as
 \begin{equation}\label{decomposation}
\Pi_{\mu\nu}(Q)=P^T_{\mu\nu}G(Q)+P^L_{\mu\nu}F(Q),
\end{equation}
here $G(Q)$ and $F(Q)$ are transverse  and longitudinal component of polarization tensor respectively. In Lorentz gauge $\partial_\mu A^\mu=0$, one  obtains
\begin{equation} \label{A0}
A_0^2=-\frac{q^2\mathbf{E}^2-q_0^2\mathbf{B}^2}{(q^2-q_0^2)^2} ,
\end{equation}
\begin{equation} \label{A}
\mathbf{A}^2=-\frac{q_0^2\mathbf{E}^2-(2q_0^2-q^2)\mathbf{B}^2}{(q^2-q_0^2)^2}.
\end{equation}
Inserting (\ref{decomposation}), (\ref{A0}) and (\ref{A}) into (\ref{action1}), one obtains
\begin{equation} \label{Gamma1}
\Gamma_1=-\frac{1}{2}\int \frac{d^4 Q}{(2\pi)^4}\Big[-\frac{F(Q)}{Q^2}\mathbf{E}^2-\Big(\frac{G(Q)}{q^2}-\frac{q_0^2 F(Q)}{q^2Q^2}\Big)\mathbf{B}^2\Big] .
\end{equation}
Combining (\ref{action0}) with (\ref{Gamma1}) and noticing the action in medium can be composed by electromagnetical field in analogy to $\Gamma_0$ which is specified as
\begin{equation} \label{action medium}
\Gamma=-\frac{1}{2}\int \frac{d^4 Q}{(2\pi)^4}(\varepsilon \mathbf{E}^2-\frac{1}{\mu}\mathbf{B}^2),
\end{equation}
we find that
\begin{equation}
\varepsilon(Q)=1-\frac{F(Q)}{Q^2} .
\end{equation}
In the static limit $q_0=0$, the dielectric function may be written as\cite{a,B}
\begin{equation}\label{dielectric function}
\varepsilon(q)=1+\frac{F(q)}{q^2},
\end{equation}
where $F({q})\equiv F(0,\mathbf{q})$.

With Born approximation, the static interaction potential is the Fourier transformation of effective propagator that involves $F$, i.e.,
\begin{equation}
U(r)=4\pi\alpha\int \frac{d^3q}{(2\pi)^3}\frac{e^{i\mathbf{q}\cdot\mathbf{r}}}{q^2\varepsilon(q) },
\end{equation}
where $\alpha=1/137$ is fine-structure constant.
Integrating over azimuth angle, one obtains\cite{a,b}
\begin{equation} \label{potential integral}
U(r)=\frac{\alpha}{\pi r}Im\int_{-\infty}^\infty dq\frac{e^{iqr}}{q\varepsilon(q)} =\frac{\alpha}{\pi r}Im\int_{-\infty}^\infty dq\frac{qe^{iqr}}{q^2+F(q)} .
\end{equation}

\section{Equation of motion in AdS/CFT}
AdS/CFT correspondence relates the $\mathcal N=4$ SYM at large $N_c$ and large 't Hooft coupling
\begin{equation} \lambda=N_cg^2_{YM}, \end{equation}
\begin{equation} 4\pi g_s=g^2_{YM}, \end{equation}
where $g_{YM}$ is the Yang-Mills coupling, $g_s$ is the string coupling, $\lambda$ is the 't Hooft coupling. For a given $\lambda$, when $N_c\to\infty$, $g_s$ goes to zero. After $\lambda\to\infty$, the quantum superstring theory reduces to classic supergravity. Thus one connects the weak coupling limit and strong coupling limit by the AdS/CFT correspondence.
The Green's function is the propagation amplitude, according to the AdS/CFT correspondence we can proceed with the formulation of the prescription for Minkowski correlators, the retarded thermal Green's function is then \cite{c}
\begin{equation} \label{Green's function}
G^R(q)=A(u)f_{-q}(u)\partial_uf_q(u)\mid_{u\to0},
\end{equation}
where $u$ is the radial coordinate, and the boundary is located at $u=0$.

Here we take the R-charge correlator of the $\mathcal N=4$ SYM as an example. Since the U(1) symmetry in R-charge model mimics the quantum electrodynamics(QED), the supersymmetric Yang-Mills interaction is considered as the imitator of QCD. Then the current-current correlator of U(1) charge from AdS/CFT involves contribution from all orders of supersymmetric Yang-Mills coupling.
According to the R-charge model presented by Huot {\it et al}\cite{d}, R-photon is a gauge field of U(1) group introduced in $\mathcal N=4$ SYM theory, which has the coupling terms in Lagrangian
\begin{equation}
 \mathcal{L}_{couple}=-\frac{1}{4}F^2_{\mu\nu}+(D_\mu\phi)^*(D^\mu\phi)-m^2\phi^*\phi,
\end{equation}
where $F_{\mu\nu}$ is the field tensor of R-photon, $D_{\mu}=\partial_\mu+ieA_\mu$ is the usual gauge-covariant derivative, and the second term gives coupling vetex of scalar field and R-photon.

In $AdS_5\times S^5$ space, and in the near-horizon limit $r\ll R$, the metric is given by
\begin{equation} \label{matric}
ds^2_{10}=\frac{(\pi TR)^2}{u}(-f(u)dt^2+dx^2+dy^2+dz^2)+\frac{R^2}{4u^2f(u)}du^2+R^2d\Omega^2_5,
\end{equation}
where $T$ is the Hawking temperature, and $f(u)=1-u^2$. The horizon corresponds to $u=1$, the spatial infinity to $u=0$. With AdS/CFT correspondence, the background (\ref{matric}) is dual to the $\mathcal N=4\ SU(N)$ SYM in the limit of $N_c\to \infty$, $g^2_{YM}N_c\to \infty$.

The five-dimensional Maxwell equation from the background (\ref{matric}) is
\begin{equation} \partial_\mu \sqrt{-g}F^{\mu\nu}=0, \end{equation}
where $\sqrt{-g}$ is the metric norm. Under the axial gauge $A_u=0$, and in the static limit $q_0=0$, the equation of motion of $A_0$ ends up with
\begin{equation}\label{EOM}
\frac{d^2A_0}{du^2}-\frac{\hat{q}^2}{u(1-u^2)}A_0=0,
\end{equation}
where $\hat{q}=q/2\pi T$.

\section{Numerical Results}
According to the Minkowski space prescription (\ref{Green's function}), the longitudinal R-photon self-energy is expressed by\cite{b}
\begin{equation} \label{Fq}
F(q)\equiv-\Pi_{00}(0,q)=-\frac{e^2N^2_cT^2}{8}\frac{A'_0(\epsilon|q)}{A_0(\epsilon|q)},
 \end{equation}
in which $N_c$ is the number of quark colors, $e^2=4\pi\alpha$, and $A_0(\epsilon|q)$ solves the equation of motion (\ref{EOM}) with $\epsilon$ the infinitesimal.

The equation (\ref{EOM}) is a Heun equation with four canonical singularities at $u=0$, $\pm 1$ and $\infty$. The indexes at the $u=1$ are 0 and 1, which give rise to two linear independent solutions
\begin{equation} \label{boundary}
A_0^{(1)}(u|q)=(1-u)+\frac{\hat{q}^2}{4}(1-u)^2+\frac{\hat{q}^2}{8}(1-u)^3+\frac{\hat{q}^4}{48}(1-u)^3+\cdots,
\end{equation}
and
\begin{equation} \label{boundary2}
A_0^{(2)}(u|q)=1+\frac{1}{2}\hat{q}^2(1-u)[\ln(1-u)+\frac{1}{2}]+\cdots.
\end{equation}
The action at horizon $u=1$ in $AdS_5$ space should be finite, so we discard the second solution eq.(\ref{boundary2}) that makes the action divergent.

We start at $1-u=\epsilon$ with  the boundary condition (\ref{boundary}), running a Range-Kutta program to solve the differential function (\ref{EOM}). Finally we obtain  $A_0$ and $A'_0$ as a function of $\hat{q}$ at $u=\epsilon$.  However we find $A'_0(\epsilon|q)$ suffers a logarithm divergence which corresponds to the ultraviolet divergence in quantum field theory. To handle this divergence, we introduce a momentum cutoff and absorb it into the renormalized coupling constant, i.e., the temperature-dependent non-divergent self-energy should be $F^s(q)=F(q)-F_0(q)$ with $F_0(q) \equiv F(q|T=0)$.

In the background (\ref{matric}), the extra dimension $u$ is related with $T$ by $u=\pi^2T^2z^2$, when $T$ goes to zero, the equation of motion becomes
\begin{equation}
\frac{d^2A_0}{du^2}-\frac{q^2}{(2\pi T)^2}\frac{A_0}{u}=0.
\end{equation}
Under the transformation $u=xT^2$,  the equation of motion of zero temperature is expressed as
\begin{equation} \frac{d^2A_0}{dx^2}-\frac{q^2}{4\pi^2}\frac{A_0}{x}=0. \end{equation}
The finite solution of the above equation of motion is
\begin{equation}
A(x|q)=\frac{q\sqrt{x}}{\pi}K_1\Big(\frac{q\sqrt{x}}{\pi}\Big),
\end{equation}
where $K_1$ is the second kind of modified Bessel function, which yields
\begin{equation}
F_0(q)=\frac{e^2N_c^2}{8}\frac{qK_0(\frac{q\sqrt{x}}{\pi})}{2\pi\sqrt{x}K_1(\frac{q\sqrt{x}}{\pi})}\bigg|_{x=\epsilon},
\end{equation}
where $K_0$ is the first kind of modified Bessel function.

In weak coupling, the $T$-dependent longitudinal photon self-energy is also calculated at one-loop level\cite{a,b}, which reads
\begin{eqnarray}
F^w(q) &=&\frac{e^2(N_c^2-1)}{2\pi^2}\Big\{\int_0^{\infty } dp\frac{p}{e^{\beta p}+1}\Big[1+\frac{p}{q}\Big(1-\frac{q^2}{4p^2}\Big)\ln\Big|\frac{2p+q}{2p-q}\Big|\Big]\nonumber\\
& & +\int_0^{\infty} dp\frac{p^2}{e^{\beta p}-1}\Big(\frac{1}{p}+\frac{1}{q}\ln\Big|\frac{2p+q}{2p-q}\Big|\Big)\Big\}.
\label{Fw}
\end{eqnarray}

\begin{figure}[h]
\begin{center}
\includegraphics[width=0.4\textwidth]{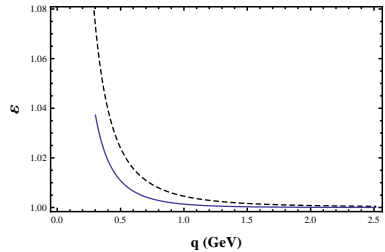}
\end{center}
\caption{Numerical results of dielectric function. The solid line is for the dielectric function in strong coupling, and the dashed line is for the dielectric function in weak coupling.}
\label{Fig.1}
\end{figure}

Inserting $F^s(q)$ and $F^w(q)$ into (\ref{dielectric function}) and setting $T=0.2$GeV and $N_c=3$, we obtain the numerical results of dielectric functions shown in Fig.\ref{Fig.1}, where solid line is for strong coupling and dashed line is for weak coupling. Obviously, the dielectric function in strong coupling is depressed compared with  that in weak coupling. Considering there is a negative correlation between the energy loss of rapid particle passing through the medium and the dielectric function of the medium\cite{energy loss}, one would expect the strong coupling system will have more energy loss than the weak coupling system.

To obtain the interaction potential (\ref{potential integral}), one could draw a contour in upper complex $q$ plane, and pick up all the residues at each pole inside the contour. Hou {\it et al} \cite{b} pointed out that the self-energy has infinite number of poles on the imaginary q-axial by using WKB approximation, which hints us to seek the poles of the integrand on the imaginary axial.

\begin{figure}[h]
\begin{center}
\includegraphics[width=0.4\textwidth]{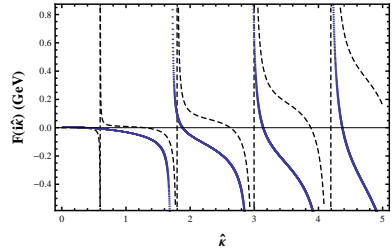}
\end{center}
\caption{Self-energy $F(i\kappa)$ for imaginary momentum. The solid line is for the numerical result of self-energy, and the dash line is for the  WKB approximation of self-energy.}
\label{Fig.2}
\end{figure}

In WKB approximation, for large and imaginary $q$, i.e.$q=i\kappa$ with $\kappa\gg T$, the renormalized self-energy reads\cite{b}
\begin{equation}
F^{\text{ WKB}}(i\kappa)\simeq-\frac{e^2N_c^2}{16\pi^2}\kappa^2\Big(\ln\frac{2}{\kappa}-\gamma_E+\frac{1}{2}\pi\tan\hat{\kappa}\delta\Big), \end{equation}
where $\gamma_E=0.5772\cdots$ is the Euler constant and
\begin{equation} \delta=\frac{1}{2\sqrt{2\pi}}\Gamma^2\Big(\frac{1}{4}\Big). \end{equation}
$F^{\text{ WKB}}(i\kappa)$ goes to infinity periodically due to the tangent function, which hints us the poles of integrand in (\ref{potential integral}), denoted by $\kappa_n$, solve the equation
\begin{equation}
\kappa^2-F(i\kappa)=0.
\end{equation}
Here $F(i\kappa)$ is the photon self-energy with imaginary argument that determined by the EOM
\begin{equation}\label{EOM2}
 \frac{d^2A_0}{du^2}+\frac{\hat{\kappa}^2}{u(1-u^2)}A_0=0
\end{equation}
by eq.(\ref{Fq}).
Constructing a contour in upper complex q plane, performing the contour integral, we can rewrite the interaction potential as
\begin{equation}\label{potential}
U(r)=2\alpha\sum_{n=1}^{\infty}\frac{1}{Res(\kappa_n)}\frac{e^{-\kappa_n r}}{r},
\end{equation}
where
\begin{equation} Res(\kappa_n)=2-\frac{1}{\kappa}\frac{\partial F(i\kappa)}{\partial\kappa}\bigg|_{\kappa=\kappa_n}.
\end{equation}
With the same Runge-Kutta program that run to solve (\ref{EOM}), we solve the  (\ref{EOM2})
and obtain the self-energy $F(i\kappa)$ as shown in Fig.\ref{Fig.2} with solid line. We also compare our numerical result with WKB approximation which is plotted in Fig.\ref{Fig.2} by dashed line.  Roughly the poles are coincident expect that the first one in our numerical result is missed in WKB approximation.

\begin{figure}[h]
\begin{center}
\includegraphics[width=0.4\textwidth]{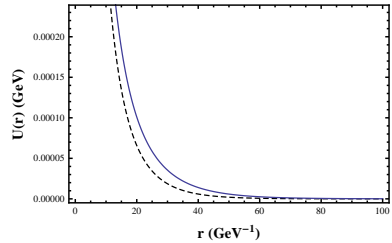}
\end{center}
\caption{Numerical result of interaction potential. The solid line is for the Debye potential in strong coupling, and the dashed line is for the Debye potential in weak coupling. }
\label{Fig.3}
\end{figure}

Since the poles appear on the exponent in (\ref{potential}), the potential (\ref{potential}) are then dominated by the first several poles. We pick up first four poles and their residues as
\begin{eqnarray}
\kappa_1=0.0641\text{GeV}, && Res(\kappa_1)=2.008,\nonumber\\
\kappa_2=2.1438\text{GeV}, &&Res(\kappa_2)=69.094,\nonumber\\
\kappa_3=3.6958\text{GeV}, &&Res(\kappa_3)=319.35,\nonumber\\
\kappa_4=5.2238\text{GeV}, &&Res(\kappa_4)=322.51,
\end{eqnarray}
and plot the potential in Fig.\ref{Fig.3} with solid line. Inserting (\ref{Fw}) into (\ref{potential integral}) and performing the integral numerically, we can also plot the interaction potential of weak coupling with dashed line in Fig.\ref{Fig.3}. It is remarkable that the strong coupling potential is over the weak coupling potential which means the strong coupling Debye mass is smaller than that in weak coupling, this result is in good agreement with the next-to-leading order Debye mass calculated by Hou {\it et al}\cite{e} and Bak {\it et al}\cite{f}. This is not surprising because the first pole in strong coupling is $\kappa_1=0.3204T$ while the Debye mass in weak coupling is

\begin{equation}
\sqrt{F^w(q\to 0)}\equiv m_D^w=\frac{eT}{2}\sqrt{N_c^2-1}\approx 0.4283T .
\end{equation}

\section{Conclusions}
In this paper, we calculated the static dielectric function and interaction potential in strong coupling limit with AdS/CFT correspondence and compared them with weak coupling ones. We found that the dielectric function in strong coupling is depressed compared with that in weak coupling, which implies more energy loss for a rapid particle passing through a strong coupling system.

After performing the contour integral in the static interaction potential in complex momentum plane, we obtained the strong coupling interaction potential. It is remarkable that the strong coupling potential is over the weak coupling one, which means the Debye mass in strong coupling is smaller than that in weak coupling. In other words, the strong coupling system present a weaker screening characteristics compared with weak coupling system.

\section*{Acknowledgements}
We thank Hai-cang Ren and De-fu Hou for help and useful advices.
This work is supported by NSFC under grant No. 10947002.

\end{document}